\documentclass[graphicx,epsf]{basi}
\begin{document}

\title[Optical spectroscopy of Classical Be stars]{Optical spectroscopy of Classical Be stars in open clusters}

\author[B. Mathew and A. Subramaniam]
       {Blesson Mathew\thanks{Current address: Astronomy and Astrophysics
           Division, Physical Research Laboratory, Navrangapura, Ahmedabad -
           380 009, Gujarat, India, email: \texttt{blesson@prl.res.in}} and 
        Annapurni Subramaniam\thanks{email: \texttt{purni@iiap.res.in}}\\
        Indian Institute of Astrophysics, Bangalore 560034, India}

\pubyear{2011}
\volume{39}
\pagerange{\pageref{firstpage}--\pageref{lastpage}}

\date{Received 2011 April 15; accepted 2011 August 17}

\maketitle

\label{firstpage}

\begin{abstract}
We present a spectroscopic study of 150 Classical Be stars in 39 open
clusters using medium resolution spectra in the wavelength range 3800 -- 9000~\AA.
One-third of the sample (48 stars in 18 clusters) has been studied for the first time. 
All these candidates were identified from an extensive survey of emission stars in young
open clusters using slitless spectroscopy (Mathew et al. 2008). This large data set
covers CBe stars of various spectral types and ages found in different cluster
environments in largely northern open clusters, and is used to study the spectral characteristics
of CBe stars in cluster environments.
About 80\% of CBe stars in our sample have H$\alpha$ equivalent width in the range $-$1 -- $-$40~\AA. 
About 86\% of the surveyed CBe stars show Fe{\sc ii} lines. 
The prominent Fe{\sc ii} lines in our surveyed stars are 4584, 5018, 5169,
5316, 6318, 6384, 7513 and 7712 \AA. 
We have identified short- and long-term line profile variability in some candidate stars 
through repeated observations. 
\end{abstract}


\section{Introduction}
A Classical Be (CBe) star is defined as a non-supergiant B-type star whose 
spectrum has, or had at some time, one or more Balmer lines in emission (Collins 1987).
The emission lines are produced in a circumstellar disc through recombination process from 
reprocessed stellar radiation. They rotate at 70$-$80\% of their critical speed and hence the 
reason for the formation of disc may not be equatorial mass loss mechanism 
(Porter \& Rivinius 2003). 
The formation and structure of circumstellar envelopes and the evolutionary status of CBe stars 
are some of the unresolved problems. Some of these problems can be tackled better through the study of 
CBe stars in open clusters since we know the age, distance and evolutionary
state of these candidates. 

McSwain \& Gies (2005) conducted a photometric survey of 55 southern open clusters
and identified 52 definite CBe stars and 129 probable
candidates. They found that the spin-up during the Terminal Age Main
Sequence (TAMS) cannot produce the
observed distribution of CBe stars while 73\% of the candidates could be spun-up by
binary mass transfer. McSwain et al. (2008) and McSwain, Huang \& Gies (2009) discovered a large number of
highly active CBe stars in NGC 3766 as well as eight other southern open clusters. 
They used H$\alpha$
spectroscopy to study the physical properties of the CBe stars in these clusters.
Using high- and medium-resolution
spectroscopy of CBe stars and binary stellar systems in young open clusters, 
Malchenko \& Tarasov (2008) found that CBe stars mostly appear at an age of 10
Myr and their concentration reaches a maximum at 12$-$20 Myr. 
Mathew, Subramaniam \& Bhatt (2008) performed a survey to identify emission stars in 
young open clusters using slitless spectroscopy. They observed 207 open star clusters and
157 emission stars were identified in 42 clusters. They found 54 new emission stars 
in 24 open clusters, of which 19 clusters were found for the first time to host these stars. 
Most of the emission stars in their survey belonged to CBe class ($\sim$92\%) while a few were
Herbig Be (HBe) stars ($\sim$6\%) and Herbig Ae (HAe) stars ($\sim$2\%). 
From the distribution of CBe stars with respect to spectral type and age, they
found that CBe stars in the spectral range B0$-$B1 have evolved into CBe phase
while others are born as CBe stars. They also found that CBe stars are present
in different evolutionary phases and hence the CBe phenomenon is unlikely to be 
only due to core contraction near the turn-off. 
From an H$\alpha$ spectroscopic survey of CBe stars in the SMC open clusters,
Martayan, Baade \& Fabregat (2009) found that certain CBe stars could be born as CBe stars
while others evolve to that phase.

Following the survey (Mathew et al. 2008), the spectra of the identified CBe stars
were obtained to study their spectral properties. The survey identified a large
number of stars covering a wide spectral and age range, thus making the sample ideal
for statistical analysis of various spectral characteristics. 
Using photometric and spectroscopic analysis we confirmed that 5 out of 
the total sample of 157 CBe stars belong to Herbig Ae/Be (HAeBe) category 
(Mathew et al. 2010). These are excluded from the present study.  
NGC 1624(1) (see Mathew et al. (2008) for nomenclature) is found to be a late O-type emission
star and hence removed from the present analysis. NGC 436(3) is also removed from
the list since the spectral features do not confirm it to be a CBe star. 
Thus, we present the spectral line details and the main results 
obtained from the spectral line analysis of 150 CBe stars. In this sample, one third of the stars (48/150) 
are studied for the first time. This is the largest sample of CBe stars present in
northern open clusters, and the data presented here is a homogeneous set. This large data set
covers CBe stars of various spectral types and ages, found in different cluster
environments of the northern open clusters. This data set is used
to identify interesting candidates for follow up observations as well as spectral variability studies. 

The paper is arranged as follows. The following section addresses the details 
of spectral observations. Section 3
explains the major results from the spectral line analysis. CBe stars which show
spectroscopic variability are addressed. 
We have also given a collective analysis of the metallic
lines in the spectra of CBe stars. The results are summarized in Section 4. 

\section{Observations}
The spectroscopic observations were done using the Himalayan Faint Object Spectrograph Camera (HFOSC) 
available with the 2.0m Himalayan Chandra Telescope (HCT), located at Hanle and 
operated by the Indian Institute
of Astrophysics (IIA). The CCD used for imaging is a 2K$\times$4K CCD,
where the central 500$\times$3500 pixels were used for spectroscopy. The pixel size is 15 $\mu$m
with an image scale of 0.297 arcsec/pixel. Slit spectra of CBe stars were taken using 
Grism 7 (3800 \AA~-- 6800 \AA) and 167 $\mu$m slit combination in the blue region which gives an effective 
resolution of 10 \AA~around the H$\beta$ line. The spectra in the red region is taken using Grism 8
(5500 \AA~-- 9000 \AA) /167 $\mu$m slit setup, which gives an effective
resolution of 7 \AA~around the H$\alpha$ line. 
The spectra were found to have good signal to noise ratio. 
The log of the observations is given in Table 1. 
All the observed spectra were wavelength calibrated and corrected
for instrument sensitivity using the Image Reduction and Analysis Facility (IRAF) tasks. 
The calibrated spectra were normalized 
and continuum fitted using IRAF tasks. 
The equivalent width (EW) of the spectral lines 
were estimated using routines in IRAF, which effectively measures the area
under the line profile. The equivalent width is measured by marking two continuum points around 
the  line  to  be  measured. The linear continuum is subtracted 
and the flux is determined by simply  summing  the  pixels  with 
partial  pixels  at  the  ends. Therefore, this method calculates the 
area under the profile irrespective of the profile shape. 
The typical error in the measurement is around 10\%. 
The telluric bands were not removed since the lines of interest were not affected by them.  

The spectral type was estimated by comparing the absorption intensities of 
higher order Balmer lines and  
He{\sc i} 4026, 4144, 4471 \AA~lines with a stellar library (Pickles 1998). 
The Balmer lines of wavelength higher than H$\delta$ are found to show filled-in 
emission features and hence not used for spectral type estimation. 
The estimated spectral types are bluer than the photometric estimates 
since the stellar flux is reddened by the cicumstellar disc (Slettebak 1985).
 
\section{Results and discussion}
The coordinates, spectral type, H$\alpha$ EW and age of the 
CBe stars are given in Table 1. The age of the CBe star corresponds to the age
of the cluster with which it is associated, as identified in Mathew et
al. (2008). From the slitless spectroscopic survey 
(Mathew et al. 2008) we identified 49 new CBe stars in 19 open clusters. 
These clusters were not known to have any CBe stars, as seen from WEBDA 
database. Among these, Bochum 6(1) was removed from the list of new CBe stars since it is 
suspected to be a HBe star (Mathew et al. 2010). 
The newly identified 48 CBe stars in 18 clusters in the improved list 
are shown in boldface in Table 1. The spectral analysis of this sample of
new CBe stars is valuable and some of these show interesting spectral features,
as shown in Table 2. The Table shows the nature of the H$\beta$ profile, presence of
metallic lines and other features. It can be noticed that the iron emission lines are present
in abundance in some stars, on the other hand they are totally absent in others.
The presence or absence of the spectral emission lines can be used to understand the
distribution of material in the circumstellar disc as a function of temperature and density. 
We plan to do a detailed study by modeling the spectral features. 
Repeated observations showed spectroscopic variability  in the case of some
candidates and these are explained below. 

\begin{table}
\caption{The journal of observations with integration time in sec, coordinates, spectral type, 
H$\alpha$ EW and age. Newly identified CBe stars are shown in boldface. 
Log of repeated observations are also given. }
\begin{tabular}{lcr cc rrr}
\hline
Emission Star & Date & Int. & RA(J2000) & Dec(J2000) & Sp.type & H$\alpha$ EW & Age \\
              &      & time & hh:mm:ss & deg:min:sec &  & \AA~ & Myr \\
\hline
{\bf Berkeley 62(1)} & 13-10-2005 & 900 & 01:01:25.82 & +63:58:25.5 & B7V & $-$14.6 & 10\\
{\bf Berkeley 63(1)} & 07-12-2005 & 600 & 02:19:32.26 & +63:43:46.4 & B2-3V & $-$31.9 & \\
                     & 01-10-2006 & 600 &             &             &       & $-$32.5 & \\
Berkeley 86(9) & 27-06-2005 & 900 & 20:20:10.75 & +38:37:30.9 & B1V & $-$5.6 & 10 \\	
Berkeley 86(26) & 27-06-2005 & 900 & 20:20:20.43 & +38:37:36.7 & B1V & $-$24.9 & 10 \\   
Berkeley 87(1) & 09-10-2005 & 900 & 20:21:59.99 & +37:26:24.1 & B1V & $-$9.5 & 8 \\    
Berkeley 87(2) & 09-10-2005 & 600 & 20:21:24.81 & +37:22:48.3 & B0-1V & $-$28.1 & 8 \\    
Berkeley 87(3) & 08-10-2005 & 900 & 20:21:28.36 & +37:26:18.9 & B2 & $-$7.9 & 8 \\    
               & 25-10-2005 & 900 &             &             &    & $-$7.4 & \\ 
Berkeley 87(4) & 09-10-2005 & 900 & 20:21:33.55 & +37:24:52.2 & B0-1V & $-$40.2 & 8 \\    
{\bf Berkeley 90(1)} & 28-08-2006 & 1200 & 20:35:41.56 & +46:46:48.9 & B0V & $-$35.3 & \\    
{\bf Bochum 2(1)} & 21-11-2005 & 600 & 06:49:07.43 & +00:21:56.3 & B5-7V & $-$18.4 & 4.6 \\  
{\bf Collinder 96(1)} & 21-11-2005 & 120 & 06:30:17.69 & +02:50:52.8 & B0-1V & $-$19.2 & 63 \\   
{\bf Collinder 96(2)} & 21-11-2005 & 180 & 06:30:30.02 & +02:53:22.0 & B5-7V & $-$5.3 & 63 \\	
IC 1590(3) & 28-09-2006 & 720 & 00:52:44.38 & +56:37:03.3 & B1V & $-$6.2 & 4 \\	
IC 4996(1) & 15-07-2005 & 600 & 20:16:29.03 & +37:38:52.3 & B3 & $-$20.5 &  8  \\   
King 10(A) & 29-07-2005 & 900 & 22:54:53.19 & +59:09:33.3 & B1V & $-$19.5 & 50 \\   
           & 30-09-2006 & 600 &             &             &     & $-$23.4 &  \\
King 10(B) & 30-07-2005 & 900 & 22:55:12.43 & +59:07:46.5 & B0V & $-$19.4 & 50 \\   
           & 01-10-2006 & 900 &             &             &     & $-$19.9 & \\
King 10(C) & 30-07-2005 & 900 & 22:55:06.47 & +59:13:10.6 & B2V & $-$12.6 & 50 \\   
           & 31-07-2005 & 900 &             &             &     & $-$12.1 &   \\
           & 01-10-2006 & 900 &             &             &     & $-$18.8 & \\
King 10(E) & 31-07-2005 & 900 & 22:54:56.64 & +59:10:22.7 & B3V & $-$15.9 & 50 \\   
           & 01-10-2006 & 900 &             &             &     & $-$11.8 & \\
{\bf King 21(B)} & 06-07-2007 & 900 & 23:49:46.84 & +62:42:35.3 & B5V & $-$18.9 &  30 \\   
{\bf King 21(C)} & 06-07-2007 & 600 & 23:49:57.83 & +62:42:07.4 & B1V & $-$6.6 & 30 \\	
{\bf King 21(D)} & 07-07-2007 & 900 & 23:49:59.04 & +62:46:21.9 & B3V & $-$16.0 & 30 \\ 
{\bf NGC 146(S1)} & 01-10-2006 & 400 & 00:32:44.72 & +63:18:15.6 & B3V & $-$21.1 & 10 \\   
{\bf NGC 146(S2)} & 01-10-2006 & 900 & 00:33:18.17 & +63:18:37.8 & B5-7V & $-$20.6 & 10 \\   
{\bf NGC 436(1)} & 09-01-2007 & 600 & 01:15:56.29 & +58:48:12.4 & B5-7V & $-$14.0 & 40 \\   
{\bf NGC 436(2)} & 10-01-2007 & 700 & 01:15:20.56 & +58:50:03.1 & B5-7V & $-$35.2 & 40 \\   
{\bf NGC 436(4)} & 09-01-2007 & 600 & 01:15:41.14 & +58:49:02.0 & B5-7V & $-$16.3 &  40 \\   
{\bf NGC 436(5)} & 09-01-2007 & 600 & 01:15:58.66 & +58:49:14.5 & B3V & $-$25.8 &  40 \\   
NGC 457(1) & 29-09-2006 & 600 & 01:19:02.36 & +58:19:20.2 & B3V & $-$16.5 &  20 \\   
NGC 457(2) & 29-09-2006 & 300 & 01:19:32.98 & +58:17:25.5 & B3V & $-$20.0 & 20 \\   
NGC 581(1) & 28-09-2006 & 300 & 01:33:41.87 & +60:42:19.4 & B2V & $-$25.6 &  12.5 \\ 
NGC 581(2) & 28-09-2006 & 600 & 01:33:24.25 & +60:39:44.9 & B2V & $-$31.6 & 12.5 \\ 
\hline
\end{tabular}
\end{table}

\begin{table}
\begin{tabular}{lrc cc rrr} 
\hline
Emission Star & Date & Int. & RA(J2000) & Dec(J2000) & Sp.type & H$\alpha$ EW & Age  \\
              &      & time & hh:mm:ss & deg:min:sec &        &  \AA~ & Myr \\
\hline
NGC 581(3) & 28-09-2006 & 120 & 01:33:15.16 & +60:41:01.7 & B0-1V & $-$15.3 &  12.5 \\ 
NGC 581(4) & 28-09-2006 & 600 & 01:33:10.96 & +60:39:30.8 & B3V & $-$14.8 & 12.5 \\ 
{\bf NGC 637(1)} & 13-10-2005 & 900 & 01:43:22.10 & +64:01:18.3 & B5-7V & $-$22.4 &  4 \\    
NGC 654(2) & 29-09-2006 & 600 & 01:44:02.89 & +61:53:18.0 & B0-1 & $-$47.9 &  10 \\    
NGC 659(1) & 21-11-2005 & 600 & 01:44:33.09 & +60:40:56.2 & B2V & $-$27.0 &  20 \\   
           & 30-09-2006 & 600 &             &             &     & $-$27.3 & \\ 
NGC 659(2) & 21-11-2005 & 600 & 01:44:28.22 & +60:40:03.4 & B1V & $-$9.9 &  20 \\   
           & 30-09-2006 & 600 &             &             &     & $-$4.1   &  \\
NGC 659(3) & 21-11-2005 & 600 & 01:44:22.80 & +60:40:43.8 & B1V & $-$14.0 &  20 \\   
           & 30-09-2006 & 600 &             &             &     & $-$13.7 & \\
NGC 663(1) & 08-10-2005 & 900 & 01:46:02.06 & +61:15:04.2 & B5V & $-$42.1 & 25 \\   
NGC 663(2) & 07-10-2005 & 300 & 01:46:06.09 & +61:13:39.1 & B0-1V & $-$33.2 & 25 \\   
NGC 663(3) & 07-10-2005 & 600 & 01:46:14.01 & +61:13:43.9 & B5V & $-$15.8 &  25 \\   
           & 09-10-2006 & 600 &             &             &     & $-$15.0  &   \\                
           & 24-10-2007 & 600 &             &             &     & $-$12.4  &   \\ 
           & 02-12-2007 & 600 &             &             &     & +1.5 & \\
NGC 663(4) & 24-10-2005 & 900 & 01:46:30.63 & +61:14:29.2 & B1V & $-$20.8 & 25 \\   
NGC 663(5) & 22-11-2005 & 600 & 01:45:46.39 & +61:09:20.9 & B1V & $-$37.7 &  25 \\   
           & 10-10-2006 & 600 &             &             &     & $-$39.8 & \\
NGC 663(6) & 24-10-2005 & 900 & 01:46:24.41 & +61:10:37.3 & B5-7V & $-$8.5 &   25 \\	
           & 24-10-2005 & 900 &             &             &       & $-$6.5 &  \\
           & 09-10-2006 & 900 &             &             &       & $-$6.8 & \\
NGC 663(7) & 24-10-2005 & 600 & 01:46:35.53 & +61:15:47.8 & B2V & $-$11.7 &  25 \\   
NGC 663(9) & 24-10-2005 & 900 & 01:46:35.60 & +61:13:39.1 & B1V & $-$54.0 &  25 \\   
NGC 663(11) & 09-10-2005 & 600 & 01:46:20.21 & +61:14:21.5 & B2V & $-$23.8 &  25 \\   
NGC 663(12) & 25-10-2005 & 600 & 01:45:37.81 & +61:07:59.1 & B2V & $-$37.1 & 25 \\   
            & 10-10-2006 & 600 &             &             &     & $-$32.7 & \\
NGC 663(12V) & 21-11-2005 & 600 & 01:46:26.84 & +61:07:41.7 & B0-1V & $-$40.8 & 25 \\   
NGC 663(13) & 22-11-2005 & 720 & 01:46:34.85 & +61:06:27.7 & B5V & -- &  25 \\	
            & 09-10-2006 & 720 &            &             &     &  $-$3.8  &   \\ 
NGC 663(14) & 25-10-2005 & 600 & 01:46:59.55 & +61:12:29.8 & B5V & $-$26.8 & 25 \\   
NGC 663(15) & 25-10-2005 & 600 & 01:47:39.34 & +61:18:20.7 & B1V & $-$42.8 & 25 \\   
NGC 663(16) & 25-10-2005 & 300 & 01:45:18.02 & +61:06:56.4 & B1V & $-$25.1 & 25 \\   
            & 10-10-2006 & 300 &             &             &     & $-$21.7 & \\
NGC 663(24) & 09-10-2005 & 600 & 01:46:28.61 & +61:13:50.4 & B3V & $-$8.0 & 25 \\	
            & 10-10-2006 & 600 &             &             &     & $-$7.9 & \\
NGC 663(P5) & 14-10-2005 & 900 & 01:45:56.11 & +61:12:45.41 & B2V & $-$30.1 & 25 \\   
            & 09-10-2006 & 900 &             &              &     & $-$25.1 & \\ 
\hline
\end{tabular}
\end{table}

\begin{table}
\begin{tabular}{lcr cc rrr} 
\hline
Emission Star & Date & Int. & RA(J2000) & Dec(J2000) & Sp.type & H$\alpha$ EW & Age  \\
              &      & time & hh:mm:ss   & deg:min:sec &        &  \AA~ & Myr \\
\hline
NGC 663(P6) & 09-10-2006 & 600 & 01:45:59.30 & +61:12:45.67 & B0-1V & $-$8.7 &  25 \\
NGC 663(P8) & 14-10-2005 & 900 & 01:45:39.63 & +61:12:59.6 & B2V & $-$21.9 & 25 \\ 
NGC 663(P23) & 25-10-2005 & 600 & 01:47:03.74 & +61:17:32.0 & B2V & $-$12.0 & 25 \\  
NGC 663(P25) & 22-11-2005 & 600 & 01:47:26.76 & +61:08:44.2 & B0-1V & $-$11.3 &  25 \\  
NGC 663(P151) & 25-10-2005 & 900 & 01:47:17.46 & +61:13:18.2 & B5-7V & $-$7.3 &  25 \\  
              & 10-10-2006 & 900 &             &             &       & $-$5.6 & \\
NGC 869(1) & 21-01-2006 & 140 & 02:19:26.65 & +57:04:42.1 & B0V & $-$67.7 & 12.5 \\
           & 24-10-2007 & 140 &             &             &     & $-$61.4 &   \\ 
           & 16-12-2007 & 140 &             &             &     & $-$60.8 & \\
NGC 869(2) & 20-01-2006 & 900 & 02:19:28.95 & +57:11:25.1 & B0-1V & $-$16.3 & 12.5 \\
           & 16-12-2007 & 900 &             &             &       & $-$15.9 & \\
NGC 869(3) & 21-01-2006 & 600 & 02:19:28.98 & +57:07:05.3 & B0-1V & $-$17.2 & 12.5 \\
           & 24-10-2007 & 600 &             &             &       & +1.7 & \\
           & 16-12-2007 & 600 &             &             &       & --  & \\
NGC 869(4) & 22-01-2006 & 300 & 02:18:47.98 & +57:04:03.0 & B0-1V & $-$5.6 &  12.5 \\
           & 24-10-2007 & 300 &            &             &     &  $-$4.9  &   \\ 
           & 16-12-2007 & 300 &            &             &     &  $-$5.2  &   \\ 
NGC 869(5) & 20-01-2006 & 600 & 02:19:13.77 & +57:07:45.0 & B1V & $-$17.6 & 12.5 \\
           & 24-10-2007 & 600 &             &             &     & $-$17.3 &  \\
           & 16-12-2007 & 600 &            &             &     &  $-$16.1 & \\
NGC 869(6) & 22-01-2006 & 900 & 02:19:08.73 & +57:03:50.0 & B5V & $-$47.8 & 12.5 \\
           & 24-10-2007 & 900 &             &             &     & $-$45.5 &  \\
           & 16-12-2007 & 900 &            &             &     &  $-$42.9 & \\
NGC 884(1) & 22-01-2006 & 240 & 02:22:48.07 & +57:12:03.6 & B0-1V & $-$16.5 & 12.5 \\
           & 15-12-2007 & 240 &            &             &     &  $-$10.0  &   \\ 
NGC 884(2) & 22-01-2006 & 180 & 02:22:06.59 & +57:05:24.6 & B0-1V & $-$70.4 &  12.5 \\
           & 15-12-2007 & 180 &             &             &       & $-$71.0 & \\
NGC 884(3) & 29-09-2006 & 300 & 02:21:52.95 & +57:09:59.3 & B0-1V & $-$6.3 &  12.5 \\
           & 15-12-2007 & 300 &             &             &       & $-$6.0 & \\
NGC 884(4) & 28-09-2006 & 480 & 02:21:44.56 & +57:10:53.9 & B2V & $-$26.8 & 12.5 \\
           & 15-12-2007 & 480 &             &             &     & $-$25.9 & \\
NGC 884(5) & 28-09-2006 & 180 & 02:21:43.39 & +57:07:31.7 & B0V & $-$12.8 &  12.5 \\
           & 15-12-2007 & 180 &             &             &     & $-$11.2 & \\
NGC 884(6) & 22-01-2006 & 600 & 02:22:02.51 & +57:09:21.1 & B1V & $-$9.9 &  12.5 \\
           & 15-12-2007 & 600 &             &             &     & $-$9.5 & \\
NGC 957(1) & 07-12-2005 & 300 & 02:33:10.45 & +57:32:52.8 & B0-1V & $-$37.9 &  10 \\  
           & 30-09-2006 & 300 &             &             &       & $-$37.8 & \\
NGC 957(2) & 07-12-2005 & 600 & 02:33:39.44 & +57:33:51.7 & B2V & $-$16.5 & 10 \\
           & 30-09-2006 & 600 &             &             &     & $-$12.7 & \\
\hline
\end{tabular}
\end{table}

\begin{table}
\begin{tabular}{lcr cc rrr} 
\hline
Emission Star & Date & Int. & RA(J2000) & Dec(J2000) & Sp.type & H$\alpha$ EW & Age \\
              &      & time & hh:mm:ss   & deg:min:sec &        &  \AA~ & Myr \\
\hline
{\bf NGC 1220(1)} & 21-11-2005 & 900 & 03:11:40.86 & +53:21:03.8 & B5V & $-$35.7 &  60 \\  
{\bf NGC 1893(1)} & 21-11-2005 & 900 & 05:22:42.95 & +33:25:05.3 & B1V & $-$69.2 & 4  \\  
{\bf NGC 2345(2)} & 07-12-2005 & 600 & 07:08:10.47 & $-$13:15:36.7 & B5V & $-$29.0 & 71 \\  
{\bf NGC 2345(5)} & 22-11-2005 & 900 & 07:08:07.53 & $-$13:13:20.7 & B5V & $-$33.8 & 71 \\  
{\bf NGC 2345(20)} & 22-11-2005 & 900 & 07:08:12.49 & $-$13:10:35.8 & B3V & $-$31.2 & 71 \\  
{\bf NGC 2345(24)} & 22-11-2005 & 900 & 07:08:11.59 & $-$13:09:27.8 & B3V & $-$27.3 &  71 \\  
                   & 15-12-2007 & 900 &             &               &     & $-$23.8 & \\
{\bf NGC 2345(27)} & 28-12-2007 & 900 & 07:08:16.09 & $-$13:10:03.6 & B3V & $-$30.9 & 71 \\  
                   & 26-02-2008 & 900 &             &               &     & $-$31.0 & \\
{\bf NGC 2345(32)} & 22-11-2005 & 900 & 07:08:19.58 & $-$13:09:41.0 & B5V & $-$19.6 & 71 \\  
                   & 09-10-2006 & 900 &             &               &     & $-$20.3 & \\
{\bf NGC 2345(35)} & 22-11-2005 & 300 & 07:08:22.84 & $-$13:10:16.5 & B2V & $-$11.2 & 71 \\  
                   & 15-12-2007 & 300 &             &               &     & $-$8.4 & \\
{\bf NGC 2345(44)} & 07-12-2005 & 600 & 07:08:25.55 & $-$13:12:01.8 & B3V & $-$11.5 &  71 \\  
                   & 09-10-2006 & 600 &             &               &     & $-$11.3 & \\
{\bf NGC 2345(59)} & 07-12-2005 & 600 & 07:08:28.04 & $-$13:15:35.8 & B3V &$-$39.0 &  71 \\  
{\bf NGC 2345(61)} & 07-12-2005 & 600 & 07:08:29.85 & $-$13:13:14.7 & B5V & $-$16.2 & 71 \\  
{\bf NGC 2345(X1)} & 07-12-2005 & 600 & 07:07:58.15 & $-$13:10:59.6 & B5V & $-$26.9 &  71 \\  
{\bf NGC 2345(X2)} & 15-12-2007 & 300 & 07:08:12.50 & $-$13:09:55.8 & B5V & $-$27.3 &  71 \\  
                   & 27-02-2008 & 300 &             &               &     & $-$24.9 & \\
{\bf NGC 2414(1)} & 07-12-2005 & 900 & 07:33:06.38 & $-$15:26:35.8 & B1V & $-$26.8 & 10 \\  
{\bf NGC 2414(2)} & 07-12-2005 & 600 & 07:33:20.44 & $-$15:27:07.0 & B1V & $-$26.1 & 10 \\  
{\bf NGC 2421(1)} & 21-01-2006 & 600 & 07:36:06.68 & $-$20:37:57.5 & B1V & $-$43.7 &  80 \\  
                  & 16-12-2007 & 600 &             &               &     & $-$34.8 & \\ 
{\bf NGC 2421(2)} & 21-01-2006 & 600 & 07:36:02.96 & $-$20:37:39.1 & B5V & $-$13.3 & 80 \\  
                  & 16-12-2007 & 600 &             &               &     & $-$11.3 & \\
{\bf NGC 2421(3)} & 16-12-2007 & 400 & 07:36:00.06 & $-$20:38:46.0 & B5V & $-$25.6 &  80 \\  
{\bf NGC 2421(4)} & 16-12-2007 & 300 & 07:36:21.95 & $-$20:37:09.6 & B5V & $-$21.8 &  80  \\ 
{\bf NGC 6649(1)} & 09-06-2007 & 300 & 18:33:28.27 & $-$10:24:07.3 & B0V & $-$35.7 & 25 \\  
                  & 06-07-2007 & 300 &             &               &     & $-$35.0 & \\
{\bf NGC 6649(2)} & 16-07-2006 & 900 & 18:33:26.16 & $-$10:23:35.9 & B5-7V & $-$15.2 &  25 \\  
                  & 09-06-2007 & 900 &             &               &       & $-$14.5 & \\
{\bf NGC 6649(3)} & 09-06-2007 & 1200 & 18:33:23.95 & $-$10:24:41.2 & B5V & $-$22.0 & 25 \\  
{\bf NGC 6649(4)} & 09-06-2007 & 1200 & 18:33:36.29 & $-$10:22:52.4 & B5V & $-$28.7 & 25 \\  
{\bf NGC 6649(5)} & 10-06-2007 & 900 & 18:33:25.34 & $-$10:20:51.5 & B5V & $-$13.2 & 25 \\  
{\bf NGC 6649(6)} & 10-06-2007 & 1200 & 18:33:34.10 & $-$10:26:04.8 & B5V & $-$24.3 & 25 \\  
{\bf NGC 6649(7)} & 10-06-2007 & 1200 & 18:33:12.32 & $-$10:25:13.5 & B2V & $-$41.1 &  25 \\  
{\bf NGC 6756(2)} & 24-10-2005 & 900 & 19:08:40.15 & +04:43:51.2 & B5V & $-$20.8 & 125 \\
\hline
\end{tabular}
\end{table}

\begin{table}
\begin{tabular}{lcr cc rrr}
\hline
Emission Star & Date  & Int. & RA(J2000) & Dec(J2000) & Sp.type & H$\alpha$ EW & Age  \\
              &       & time & hh:mm:ss   & deg:min:sec &    &  \AA~ & Myr \\
\hline  
{\bf NGC 6756(3)} & 24-10-2005 & 900 & 19:08:46.24 & +04:40:23.2 & B5V & $-$7.5 & 125 \\  
NGC 6834(1) & 07-10-2005 & 900 & 19:52:06.48 & +29:24:37.7 & B5V & $-$38.9 & 40 \\  
            & 05-07-2007 & 900 &             &             &     & $-$37.5 & \\
NGC 6834(2) & 07-10-2005 & 600 & 19:52:09.53 & +29:23:34.0 & B1V & $-$42.5 & 40 \\  
            & 06-07-2007 & 600 &             &             &     & $-$41.7 & \\
NGC 6834(3) & 07-10-2005 & 600 & 19:52:21.21 & +29:20:20.4 & B5V & $-$14.5 & 40 \\  
            & 07-07-2007 & 600 &             &             &     & $-$12.8 &  \\
NGC 6834(4) & 07-10-2005 & 600 & 19:52:16.62 & +29:25:15.0 & B3V & $-$9.2 &  40 \\  
{\bf NGC 6910(A)} & 29-07-2005 & 900 & 20:23:11.74 & +40:43:25.9 & B3V & $-$36.2 & 6.3 \\ 
{\bf NGC 6910(B)} & 29-07-2005 & 300 & 20:23:09.74 & +40:45:53.0 & B3V & $-$10.5 &  6.3 \\ 
{\bf NGC 7039(1)} & 21-11-2005 & 900 & 21:11:00.95 & +45:39:41.4 & B1-3V & $-$46.6 & 1000 \\ 
NGC 7128(1) & 14-10-2005 & 900 & 21:44:02.92 & +53:42:12.4 & B1V & $-$43.6 & 10  \\ 
NGC 7128(2) & 14-10-2005 & 900 & 21:44:05.25 & +53:42:36.8 & B5V & $-$8.1 & 10  \\ 
NGC 7128(3) & 14-10-2005 & 900 & 21:43:33.57 & +53:45:32.3 & B1V & $-$18.9 &  10  \\ 
NGC 7235(1) & 14-10-2005 & 900 & 22:12:19.54 & +57:16:04.1 & B0-1V & $-$36.2 &12.5 \\
            & 25-10-2005 & 900 &             &             &       & $-$39.4 & \\ 
NGC 7261(1) & 07-12-2005 & 600 & 22:19:51.44 & +58:08:53.5 & B0V & $-$48.1 & 46 \\  
NGC 7261(2) & 07-12-2005 & 600 & 22:20:10.09 & +58:06:34.3 & B1V & $-$14.4 & 46 \\  
NGC 7261(3) & 07-12-2005 & 600 & 22:20:13.31 & +58:07:45.5 & B0-1V & $-$38.8 &  46 \\ 
NGC 7380(1) & 17-07-2006 & 900 & 22:47:42.62 & +58:07:46.8 & B5-7V & $-$28.0 &  10 \\  
            & 01-08-2007 & 900 &             &             &       & $-$28.6 & \\ 
NGC 7380(2) & 17-07-2006 & 600 & 22:47:40.12 & +58:09:03.7 & B1-3V & $-$36.7 &  10 \\  
NGC 7380(3) & 17-07-2006 & 300 & 22:47:49.56 & +58:08:49.6 & B1-3V & $-$21.9 &  10 \\  
NGC 7419(A) & 15-07-2005 & 900 & 22:54:36.68 & +60:48:35.2 & B0V & $-$45.3 & 25 \\  
NGC 7419(B) & 27-06-2005 & 900 & 22:54:27.12 & +60:48:52.2 & B0V & $-$52.1 &  25 \\  
NGC 7419(C) & 31-07-2005 & 900 & 22:54:25.62 & +60:49:01.4 & B1V & $-$7.0 &  25 \\  
NGC 7419(D) & 15-07-2005 & 900 & 22:54:23.76 & +60:49:31.0 & B1V & $-$41.2 & 25 \\  
NGC 7419(E) & 31-07-2005 & 900 & 22:54:24.36 & +60:47:36.2 & B6 & $-$62.1 &  25 \\  
            & 09-10-2005 & 900 &             &             &    & $-$55.2 & \\
NGC 7419(F) & 21-01-2006 & 900 & 22:54:24.28 & +60:47:01.6 & B0V & $-$39.3 &  25 \\  
            & 01-12-2008 & 900 &             &             &     & $-$41.8 & \\
NGC 7419(G) & 08-08-2005 & 900 & 22:54:20.49 & +60:49:52.5 & B0V & $-$51.6 &  25 \\  
NGC 7419(H) & 15-07-2005 & 1200 & 22:54:19.54 & +60:48:52.0 & B1-3V & $-$9.9 & 25 \\  
            & 08-08-2005 & 1200 &            &             &     &  $-$8.6   &   \\ 
            & 09-10-2006 & 1200 &            &             &     &  $-$3.4   &   \\ 
NGC 7419(I) & 15-07-2005 & 1200 & 22:54:19.65 & +60:48:35.8 & B0V & $-$20.6 &  25 \\  
NGC 7419(I1) & 09-10-2005 & 900 & 22:53:53.24 & +60:48:08.3 & B1-3V & $-$20.6 & 25 \\  
NGC 7419(J) & 08-08-2005 & 900 & 22:54:15.34 & +60:49:49.9 & B0V & $-$15.3 & 25 \\  
\hline
\end{tabular}
\end{table}

\begin{table}
\begin{tabular}{lcr cc rrr}
\hline
Emission Star & Date  & Int. & RA(J2000) & Dec(J2000) & Sp.type & H$\alpha$ EW & Age \\
              &       & time & hh:mm:ss   & deg:min:sec &    &  \AA~ & Myr \\
\hline   
NGC 7419(K) & 21-01-2006 & 900 & 22:54:20.47 & +60:48:53.9 & B1V & $-$1.2 &  25 \\  
            & 09-10-2006 & 900 &            &             &     &  $-$1.5   &   \\ 
NGC 7419(L) & 15-07-2005 & 1200 & 22:54:17.86 & +60:48:57.2 & B0V & $-$52.3 & 25 \\  
NGC 7419(M) & 08-08-2005 & 900 & 22:54:14.55 & +60:48:39.1 & B0V & $-$34.1 & 25 \\  
NGC 7419(N) & 08-08-2005 & 900 & 22:54:15.90 & +60:47:49.2 & B2.5 & $-$61.0 & 25 \\ 
            & 09-10-2006 & 900 &            &             &     &  $-$57.4 & \\
NGC 7419(O) & 08-08-2005 & 900 & 22:54:07.03 & +60:48:18.0 & B8V & $-$32.4 & 25 \\  
            & 09-10-2006 & 900 &            &             &     &  $-$27.9 & \\
NGC 7419(P) & 08-10-2005 & 900 & 22:54:13.97 & +60:46:20.4 & B0V & $-$19.5 & 25 \\  
            & 10-10-2006 & 900 &            &             &     &  $-$22.8  &   \\ 
NGC 7419(Q) & 08-10-2005 & 900 & 22:54:14.83 & +60:51:22.7 & B1-3V & $-$64.3 & 25 \\  
NGC 7419(R) & 08-10-2005 & 900 & 22:54:17.07 & +60:51:37.6 & B4V & $-$48.3 & 25 \\  
            & 10-10-2006 & 900 &             &             &     & $-$39.3 & \\ 
NGC 7419(1) & 07-10-2005 & 900 & 22:54:29.22 & +60:49:08.0 & B0V & $-$51.4 & 25 \\  
            & 09-10-2006 & 900 &             &             &     & $-$48.4 & \\
NGC 7419(2) & 07-10-2005 & 900 & 22:54:26.46 & +60:49:06.3 & B1V & $-$21.1 & 25 \\  
NGC 7419(3) & 08-08-2005 & 900 & 22:54:22.56 & +60:49:53.1 & B0V & $-$49.5 & 25 \\  
            & 09-10-2006 & 900 &             &             &     & $-$53.7 & \\
NGC 7419(4) & 07-10-2005 & 900 & 22:54:23.00 & +60:50:04.4 & B0V & $-$34.3 & 25 \\  
NGC 7419(5) & 07-10-2005 & 900 & 22:54:07.58 & +60:50:22.9 & B0V & $-$62.3 & 25 \\  
NGC 7419(6) & 08-10-2005 & 900 & 22:54:26.05 & +60:47:57.1 & B5 & $-$37.5 & 25 \\  
            & 09-10-2006 & 900 &             &             &    & $-$33.7 & \\
NGC 7510(1A) & 24-10-2005 & 900 & 23:10:57.76 & +60:33:57.1 & B0V & $-$36.1 & 10 \\  
NGC 7510(1B) & 12-10-2005 & 900 & 23:11:08.53 & +60:35:03.9 & B1V & $-$23.8 & 10 \\  
NGC 7510(1C) & 13-10-2005 & 600 & 23:10:47.75 & +60:31:52.7 & B0V & $-$58.1 & 10 \\  
Roslund 4(1) & 25-10-2005 & 900 & 20:04:50.44 & +29:11:06.1 & B3V & $-$37.7 & 16 \\  
             & 03-12-2008 & 900 &             &             &     & $-$48.0 & \\
Roslund 4(2) & 25-10-2005 & 120 & 20:04:47.07 & +29:10:03.2 & B0V & $-$62.6 & 16 \\ 
\hline
\end{tabular}
\end{table}

\begin{table}
\caption{Spectral lines identified in CBe stars}
\begin{tabular}{lrrrr}
\hline
CBe star & Nature of H$\beta$ & Fe{\sc ii} & O{\sc i} & Other features\\
\hline
Berkeley 62(1) & H$\beta$(a) & none & 8446(e) & -- \\
Berkeley 63(1) & H$\beta$(eina) & 4e & 7772(a), 8446(e) & -- \\
Berkeley 86(9) & H$\beta$(a) & 2a, 1e & 7772(a), 8446(e) & -- \\ 
Berkeley 86(26) & H$\beta$(e) & 4a, 3e & 7772(a), 8446(e) & -- \\
Berkeley 87(1) & H$\beta$(eina) & none & 8446(e) & -- \\
Berkeley 87(2) & H$\beta$(e) & 2e,1a & 8446(e) & -- \\
Berkeley 87(3) & H$\beta$(a) & none & 7772(e), 8446(e) & --  \\
Berkeley 87(4) & H$\beta$(e) & 7e,1a & 7772(e), 8446(e) & -- \\
Berkeley 90(1) & H$\beta$(e) & 7e & 7772(e), 8446(e) & -- \\
Bochum 2(1) & H$\beta$(fill-in) & 1(a) & 7772(a), 8446(e) & 6347, 6371(Si{\sc ii},a)\\
Collinder 96(1) & H$\beta$(e) & 4e & 7772(e), 8446(e) & -- \\
Collinder 96(2) & H$\beta$(a) & none & 7772(a) & -- \\
IC 1590(3) & H$\beta$(a) & none & 7772(a) & -- \\
IC 4996(1) & H$\beta$(a) & 1e & 8446(e) & -- \\
King 10(A) & H$\beta$(e) & 2e,1a & 8446(e) & 6347(Si{\sc ii},a)\\
King 10(B) & H$\beta$(e) & 1e & 8446(e) & -- \\
King 10(C) & H$\beta$(eina) & 1e & 8446(e) & -- \\
King 10(E) & H$\beta$(eina) & 1e & 8446(e) & -- \\
King 21(B) & H$\beta$(eina) & 1e & 7772(a), 8446(e) & -- \\
King 21(C) & H$\beta$(a) & none & 8446(e) & -- \\
King 21(D) & H$\beta$(eina) & none & 8446(e) & -- \\
NGC 146(S1) & H$\beta$(eina) & 11a,1e & 7772(a), 8446(e) & 6347, 6371(Si{\sc ii},a)\\
NGC 146(S2) & H$\beta$(eina) & none & 8446(e) & -- \\
NGC 436(1) & H$\beta$(a) & none & 7772(a), 8446(e) & 6347, 6371(Si{\sc ii},a)\\
NGC 436(2) & H$\beta$(eina) & 9e & 7772(e), 8446(e) & -- \\
NGC 436(4) & H$\beta$(a) & 3a,1e & 7772(a), 8446(e) & 6347, 6371(Si{\sc ii},a)\\
NGC 436(5) & H$\beta$(eina) & 8e & 7772(e), 8446(e) & -- \\
NGC 457(1) & H$\beta$(eina) & none & 8446(e) & -- \\
NGC 457(2) & H$\beta$(eina) & 8a,2e & 7772(a), 8446(e) & 6347, 6371(Si{\sc ii},a)\\
NGC 581(1) & H$\beta$(eina) & 2e & 7772(e), 8446(e) & -- \\
NGC 581(2) & H$\beta$(eina) & 4a,3e & 7772(a), 8446(e) & 6347, 6371(Si{\sc ii},a)\\
NGC 581(3) & H$\beta$(eina) & 4e,1a & 7772(e), 8446(e) & -- \\
NGC 581(4) & H$\beta$(a) & none & 7772(a), 8446(e) & 6347, 6371(Si{\sc ii},a)\\
NGC 637(1) & H$\beta$(eina) & none & 8446(e) & -- \\
NGC 654(2) & H$\beta$(e) & 10e,2a & 7772(e), 8446(e) & -- \\
NGC 659(1) & H$\beta$(eina) & 2dpe & 7772(dpe), 8446(e) & -- \\
NGC 659(2) & H$\beta$(a) & none & 7772(a), 8446(a) & -- \\
NGC 659(3) & H$\beta$(eina) & none & 7772(e), 8446(e) & -- \\
NGC 663(1) & H$\beta$(eina) & 2a,1e & 7772(e), 8446(e) & --  \\
NGC 663(2) & H$\beta$(e) & 5e,5a & 7772(e), 8446(e) & --  \\  
NGC 663(3) & H$\beta$(fill-in) & 5a,2e & 7772(a), 8446(e) &  6347, 6371(Si{\sc ii},a)\\
NGC 663(4) & H$\beta$(eina) & 5e & 7772(e), 8446(e) & 7896(Mg{\sc ii},e)\\ 
\hline
\end{tabular}
\\ e - emission profile, a - absorption, dpe - double-peaked emission\\
eina - emission in absorption, ce - core-emission\\
\end{table}

\begin{table}
\begin{tabular}{lrrrr}
\hline
CBe star & Nature of H$\beta$ & Fe{\sc ii} & O{\sc i} & Other features\\
\hline
NGC 663(5) & H$\beta$(eina) & 10e & 7772(e), 8446(e) & 6347(Si{\sc ii},e) \\
           &                 &      &                 & 7896(Mg{\sc ii},e)\\
NGC 663(6) & H$\beta$(a) & 3a & none & 6371(Si{\sc ii},a)\\
NGC 663(7) & H$\beta$(a) & 1e & 7772(a), 8446(e) & 6347(Si{\sc ii},a)\\
NGC 663(9) & H$\beta$(e) & 15e & 7772(e), 8446(e) & -- \\
NGC 663(11) & H$\beta$(eina) & 5e & 7772(e), 8446(e) & --  \\
NGC 663(12) & H$\beta$(eina) & 7e & 8446(e) & --  \\
NGC 663(12V) & H$\beta$(e) & 12e & 7772(e), 8446(e) & -- \\
NGC 663(13) & H$\beta$(eina) & 3a,1e & 7772(a), 8446(e) & 6347, 6371(Si{\sc ii},a)\\
NGC 663(14) & H$\beta$(eina) & 4e,2a & 7772(e), 8446(e) & -- \\
NGC 663(15) & H$\beta$(e) & 12e & 7772(e), 8446(e) & -- \\ 
NGC 663(16) & H$\beta$(e) & 1e & 8446(e) & -- \\
NGC 663(24) & H$\beta$(a) & 1a,1e & 7772(a) & --  \\
NGC 663(P5) & H$\beta$(eina) & 7e & 7772(e), 8446(e) & 7896(Mg{\sc ii},e)\\
NGC 663(P6) & H$\beta$(fill-in) & 1e & 8446(e) & -- \\
NGC 663(P8) & H$\beta$(eina) & 2a,2e & 8446(e) & --  \\
NGC 663(P23) & H$\beta$(fill-in) & 6e & 7772(e), 8446(e) & 6371(Si{\sc ii},a)\\
NGC 663(P25) & H$\beta$(eina) & 5e & 7772(a), 8446(e) & 6371(Si{\sc ii},a)\\
NGC 663(P151) & H$\beta$(a) & 3a & 7772(a), 8446(e) & 6371(Si{\sc ii},a)\\
NGC 869(1) & H$\beta$(e) & 21e & 7772(e), 8446(e) & 6347(Si{\sc ii},e) \\
           &              &     &                  & 7896(Mg{\sc ii},e)\\
NGC 869(2) & H$\beta$(eina) & 7e,1a & 7772(e), 8446(e) & -- \\
NGC 869(3) & H$\beta$(a) & 2e & 7772(e), 8446(e) & -- \\
NGC 869(4) & H$\beta$(a) & 5e & 7772(a), 8446(e) & -- \\
NGC 869(5) & H$\beta$(eina) & 9e & 7772(e), 8446(e) & --  \\
NGC 869(6) & H$\beta$(e) & 6a,3e & 7772(a), 8446(e) & 6347, 6371(Si{\sc ii},a)\\
NGC 884(1) & H$\beta$(eina) & none & 7772(e), 8446(e) & -- \\
NGC 884(2) & H$\beta$(e) & 24e & 7772(e), 8446(e) & 6347, 6371(Si{\sc ii},e) \\
           &              &     &                  & 7896(Mg{\sc ii},e)\\
NGC 884(3) & H$\beta$(fill-in) & 3e & 7772(e), 8446(e) & --\\
NGC 884(4) & H$\beta$(fill-in) & 6e & 7772(e), 8446(e) & --\\
NGC 884(5) & H$\beta$(e) & 1e & 7772(e), 8446(e) & 6347, 6371(Si{\sc ii},e)\\
NGC 884(6) & H$\beta$(a) & 6e,2a & 7772(a), 8446(e) & 6347, 6371(Si{\sc ii},a)\\
NGC 957(1) & H$\beta$(e) & 9e & 7772(e), 8446(e) & 6347, 6371(Si{\sc ii},e) \\
           &              &    &                  & 7896(Mg{\sc ii})\\
NGC 957(2) & H$\beta$(eina) & 6e,1a & 7772(e), 8446(e) & 7896(Mg{\sc ii})\\
NGC 1220(1) & H$\beta$(eina) & 2e & 7772(e), 8446(e) & -- \\
NGC 1893(1) & H$\beta$(e) & 19e & 7772(e), 8446(e) & 6347(Si{\sc ii},e)\\
NGC 2345(2) & H$\beta$(fill-in) & 2e & 7772(a), 8446(e) & 6347, 6371(Si{\sc ii},a)\\
NGC 2345(5) & H$\beta$(eina) & 5e & 8446(e) & 6347(Si{\sc ii},a)\\
NGC 2345(20) & H$\beta$(eina) & 3e & 8446(e) & --  \\
NGC 2345(24) & H$\beta$(eina) & none & 8446(e) & --  \\
NGC 2345(27) & H$\beta$(eina) & 14e & 7772(e), 8446(e) & 6347(Si{\sc ii},a)\\
NGC 2345(32) & H$\beta$(fill-in) & none & 8446(e) & 6347, 6371(Si{\sc ii},a)\\
NGC 2345(35) & H$\beta$(a) & none & 8446(e) & 6347, 6371(Si{\sc ii},a)\\
\hline
\end{tabular}
\end{table}

\begin{table}
\begin{tabular}{lrrrr}
\hline
CBe star & Nature of H$\beta$ & Fe{\sc ii} & O{\sc i} & Other features\\
\hline
NGC 2345(44) & H$\beta$(fill-in) & none & 7772(a) & 6371(Si{\sc ii},a)\\
NGC 2345(59) & H$\beta$(eina) & 4e & 7772(e), 8446(e) & -- \\
NGC 2345(61) & H$\beta$(a) & none & 7772(a), 8446(e) & 6347(Si{\sc ii},a)\\
NGC 2345(X1) & H$\beta$(eina) & 2e & 8446(e) & 6371(Si{\sc ii},a)\\
NGC 2345(X2) & H$\beta$(eina) & 5e & 7772(e), 8446(e) & -- \\ 
NGC 2414(1) & H$\beta$(eina) & 4e & 7772(a), 8446(e) & --  \\
NGC 2414(2) & H$\beta$(eina) & 2e & 7772(e), 8446(e) & -- \\
NGC 2421(1) & H$\beta$(eina) & 12e & 7772(e), 8446(e) & -- \\
NGC 2421(2) & H$\beta$(a) & none &  8446(e) & -- \\
NGC 2421(3) & H$\beta$(eina) & 4e & 7772(e), 8446(e) & -- \\
NGC 2421(4) & H$\beta$(eina) & 5e & 7772(e), 8446(e) & -- \\
NGC 6649(1) & H$\beta$(e) & 11e & 7772(e), 8446(e) & -- \\
NGC 6649(2) & H$\beta$(fill-in) & 5e & 7772(a), 8446(e) & -- \\
NGC 6649(3) & H$\beta$(fill-in) & 4e & 7772(e), 8446(e) & 6371(Si{\sc ii},a)\\
NGC 6649(4) & H$\beta$(fill-in) & 5e & 7772(e), 8446(e) & 6347, 6371(Si{\sc ii},a)\\
NGC 6649(5) & H$\beta$(fill-in) & 2e & 8446(e) & 6371(Si{\sc ii},a)\\
NGC 6649(6) & H$\beta$(fill-in) & 5e & 8446(e) & 6371(Si{\sc ii},a)\\
NGC 6649(7) & H$\beta$(eina) & 14e & 7772(e), 8446(e) & 6347, 6371(Si{\sc ii},a)\\
NGC 6756(2) & H$\beta$(eina) & 4e,1a & 8446(e) & -- \\
NGC 6756(3) & H$\beta$(a) & 1e,1a & 8446(e) & -- \\
NGC 6834(1) & H$\beta$(eina) & 18e & 7772(e), 8446(e) & 5463(N{\sc ii},e)\\
NGC 6834(2) & H$\beta$(eina) & 18e,2a & 7772(e), 8446(e) & -- \\
NGC 6834(3) & H$\beta$(a) & 5e & 8446(e) &  -- \\
NGC 6834(4) & H$\beta$(a) & 2e & 7772(a), 8446(e) & -- \\
NGC 6910(A) & H$\beta$(eina) & 5e,4a & 7772(a), 8446(e) & 7896(Mg{\sc ii},e) \\
            &                 &       &                  & 6347, 6371(Si{\sc ii},a)\\
NGC 6910(B) & H$\beta$(a) & 2e & none & --  \\
NGC 7039(1) & H$\beta$(eina) & 16e & 7772(e), 8446(e) & -- \\
NGC 7128(1) & H$\beta$(eina) & 14e & 7772(e), 8446(e) & -- \\
NGC 7128(2) & H$\beta$(a) & 2a & 7772(a), 8446(e) & 6347, 6371(Si{\sc ii},a)\\
NGC 7128(3) & H$\beta$(eina) & 3a,1e & 8446(e) & --  \\ 
NGC 7235(1) & H$\beta$(ce) & 16e & 7772(e), 8446(e) & --  \\
NGC 7261(1) & H$\beta$(ce) & 13e & 7772(e), 8446(e) & 7896(Mg{\sc ii},e) \\
            &               &     &                  & 6347, 6371(Si{\sc ii},e)\\
NGC 7261(2) & H$\beta$(eina) & 1e & 8446(e) & --  \\
NGC 7261(3) & H$\beta$(e) & 13e & 7772(e), 8446(e) & 7896(Mg{\sc ii},e) \\ 
            &              &     &                  & 6347, 6371(Si{\sc ii},e)\\
NGC 7380(1) & H$\beta$(eina) & none & 7772(a), 8446(e) & 6347, 6371(Si{\sc ii},a)\\
NGC 7380(2) & H$\beta$(e) & 5e & 7772(e), 8446(e) & --  \\
NGC 7380(3) & H$\beta$(eina) & 11e & 7772(e), 8446(e) & 5942(N{\sc ii},e)\\
NGC 7419(A) & H$\beta$(e) & 13e & 7772(e), 8446(e) & 6371(Si{\sc ii},e)\\
NGC 7419(B) & H$\beta$(e) & 10e & 7772(e), 8446(e) & -- \\
NGC 7419(C) & H$\beta$(a) & 1e & 7772(a), 8446(e) & 5005(N{\sc ii},e)\\
NGC 7419(D) & H$\beta$(eina) & 6e & 7772(e), 8446(e) & 6347(Si{\sc ii},e) \\
            &                 &    &                  & 7877, 7896(Mg{\sc ii},e)\\
\hline
\end{tabular}
\end{table}

\begin{table}
\begin{tabular}{lrrrr}
\hline
CBe star & Nature of H$\beta$ & Fe{\sc ii} & O{\sc i} & Other features\\
\hline
NGC 7419(E) & H$\beta$(e) & 4e & 7772(e), 8446(e) & -- \\ 
NGC 7419(F) & H$\beta$(a) & 1e & 8446(e) &  -- \\
NGC 7419(G) & H$\beta$(e) & 11e & 7772(e), 8446(e) & -- \\ 
NGC 7419(H) & H$\beta$(a) & 4e & 7772(a), 8446(e) &  --\\
NGC 7419(I) & H$\beta$(eina) & 9e & 7772(e), 8446(e) & 4131(Si{\sc ii},a)\\
NGC 7419(I1) & H$\beta$(eina) & none & 7772(a), 8446(e) & -- \\
NGC 7419(J) & H$\beta$(eina) & 5e & 7772(e), 8446(e) & -- \\
NGC 7419(K) & H$\beta$(a) & 4e & 7772(a), 8446(e) & -- \\
NGC 7419(L) & H$\beta$(e) & 15e & 7772(e), 8446(e) & 6347(Si{\sc ii},e), 5530(N{\sc ii},e)\\
NGC 7419(M) & H$\beta$(e) & 14e & 7772(e), 8446(e) & 6371(Si{\sc ii},a)\\
NGC 7419(N) & H$\beta$(e) & 12e & 7772(e), 8446(e) & 7896(Mg{\sc ii},a)\\
NGC 7419(O) & H$\beta$(a) & 3e & 7772(e), 8446(e) & 7877(Mg{\sc ii},e)\\
NGC 7419(P) & H$\beta$(a) & 5e & 7772(a), 8446(e) & 5684(N{\sc ii},e) \\
            &              &    &                   & 6371(Si{\sc ii},a), 7896(Mg{\sc ii},a)\\
NGC 7419(Q) & H$\beta$(e) & 12e,1a & 7772(e), 8446(e) &  6347(Si{\sc ii},a),7877(Mg{\sc ii},a)\\
NGC 7419(R) & H$\beta$(e) & 5e & 7772(a), 8446(e) & -- \\
NGC 7419(1) & H$\beta$(e) & 5e & 7772(e), 8446(e) & 7896(Mg{\sc ii},a)\\
NGC 7419(2) & H$\beta$(a) & 2e & 7772(a), 8446(e) & -- \\
NGC 7419(3) & H$\beta$(e) & 9e & 7772(e), 8446(e) & 7896(Mg{\sc ii},a)\\
NGC 7419(4) & H$\beta$(e) & 7e & 8446(e) & -- \\
NGC 7419(5) & H$\beta$(e) & 6e,1a & 7772(e), 8446(e) & 5711(N{\sc ii},e), 7896(Mg{\sc ii},a)\\
            &              &       &                  & 6347, 6371(Si{\sc ii},e), \\
NGC 7419(6) & H$\beta$(e) & 8e & 8446(e) & 6347(Si{\sc ii},a)\\
NGC 7510(A) & H$\beta$(e) & 7e,3a & 7772(a), 8446(e) & -- \\
NGC 7510(B) & H$\beta$(eina) & 5e,1a & 7772(a), 8446(e) & -- \\
NGC 7510(C) & H$\beta$(e) & 12e & 7772(a), 8446(e) & 7877, 7896(Mg{\sc ii},e) \\
            &              &     &                  & 6347, 6371(Si{\sc ii},e)\\
Roslund 4(1) & H$\beta$(a) & 1e,2a & 7772(a), 8446(e) & -- \\
Roslund 4(2) & H$\beta$(e) & 17e,1a & 7772(a), 8446(e) & 6347, 6371(Si{\sc ii},a)\\
             &              &        &                  & 7896(Mg{\sc ii},e)\\
\hline
\end{tabular}
\end{table}

\subsection{Spectroscopic variability}
CBe stars are found to show long- and short-term variability in spectral
features. The long-term variability can be due to activity in the
circumstellar disc. To understand the variability in spectral features we took
multiple spectra of most of the CBe stars and those which show major changes are
given below. 
The candidates which only showed variability in H$\alpha$ line profile 
are discussed in the next subsection. 

{\bf NGC 659(2)}: The H$\alpha$ profile changed from normal state when
observed on 21-11-2005 to core emission on 30-09-2006 with a significant reduction in emission
strength, $-$10 to $-$4.1 \AA~ (see Fig. 1). 
O{\sc i} 7772 \AA~(unresolved triplet of 7772, 7774, 7775), 
Ca{\sc ii} triplet and Paschen absorption lines were absent/structured during
initial observations while O{\sc i} 8446 \AA~was seen in emission. 
After a period of 10 months it can be seen that these lines disappeared from the spectra. 
Paschen, He{\sc i} and all the Balmer lines other than H$\alpha$ are present in absorption,
with no visible emission component. 
The circumstellar disc was getting dissipated during the period of the observations. 

\begin{figure}
\centerline{\includegraphics[width=9cm]{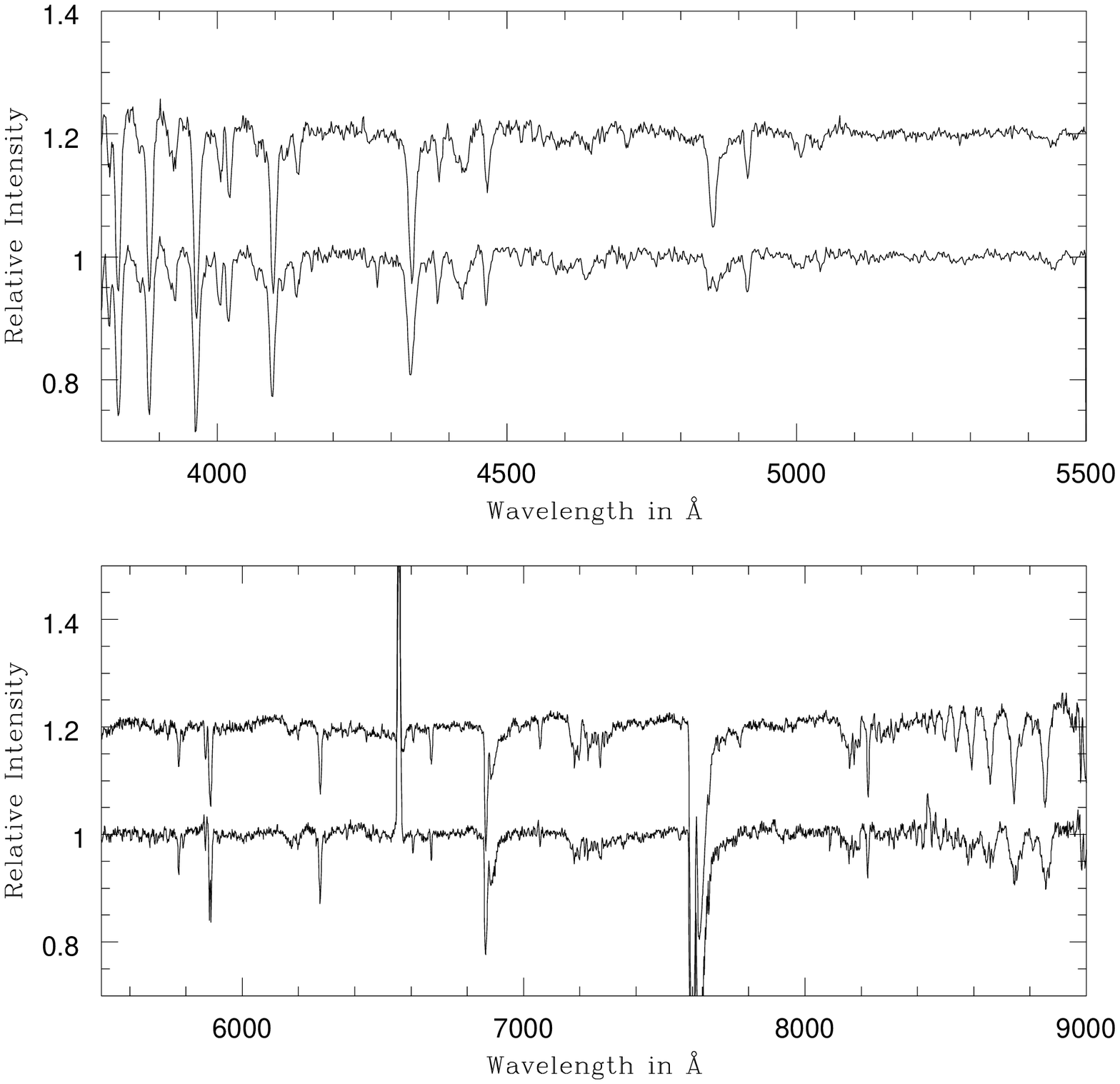}}
\caption{Spectra of CBe star NGC 659(2) in the wavelength range 3800 -- 9000~\AA. 
The spectra are from observations done on 21-11-2005 (lower) and 30-09-2006 (upper).}
\end{figure}

\begin{figure}
\centerline{\includegraphics[width=9cm]{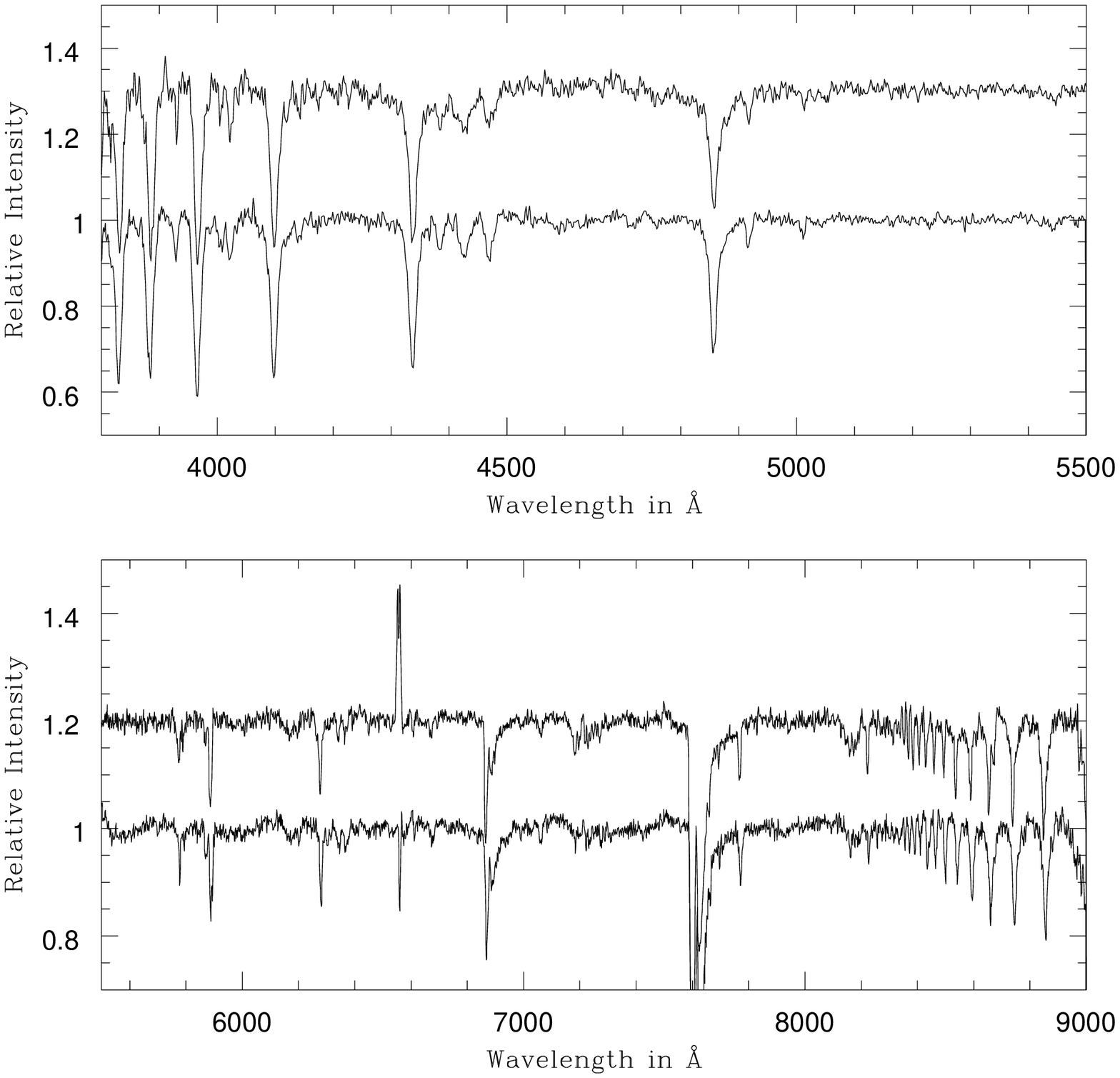}}
\caption{Spectra of CBe star NGC 663(13) in the wavelength range 3800 -- 9000
  \AA. The spectra are from observations done on 22-11-2005 (lower) and 09-10-2006 (upper).}
\end{figure}

\begin{figure}
\centerline{\includegraphics[width=9cm]{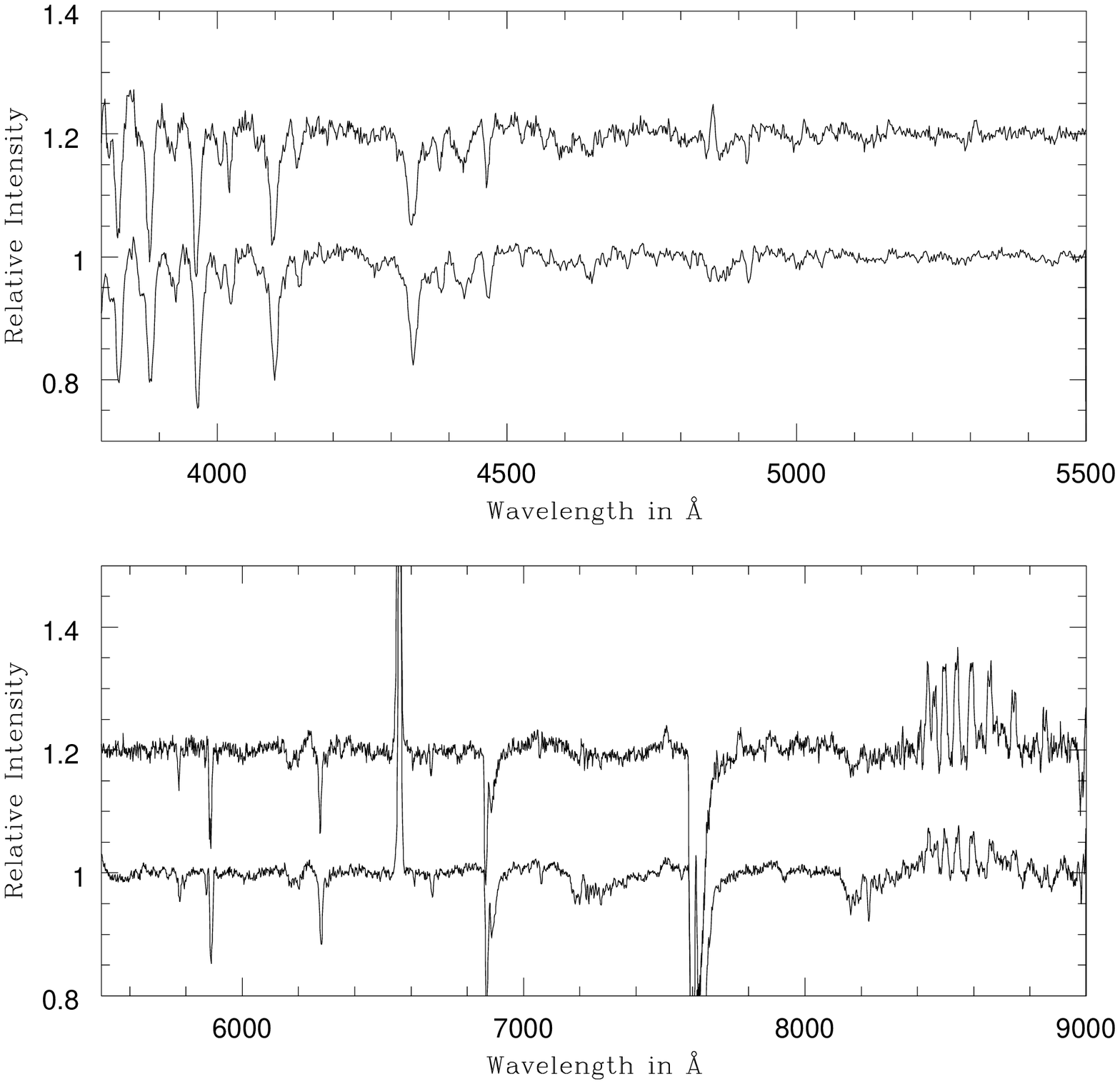}}
\caption{Spectral variability in CBe star NGC 884(1) when observed on
  22-01-2006 (lower spectra) and 15-12-2007 (upper spectra).}
\end{figure}

{\bf NGC 663(13)}: The H$\alpha$ profile was in 
absorption when observed on 22-11-2005 which changed to a double-peaked emission
profile on 09-10-2006 (Fig. 2). 
This is associated with decrease in absorption strength for other
Balmer lines. The Paschen lines were found to be unchanged while Ca{\sc ii} lines
got intensified, as identified from the deepening of absorption component in 
Ca{\sc ii} 8662 \AA~and P13 line complex. 
Hence we have identified the formation of a circumstellar disc in NGC
663(13) over a period of 1 year, from which emission lines of 
H$\alpha$ and Ca{\sc ii} are formed. 

{\bf NGC 884(1)}: The H$\alpha$ emission strength appear enhanced in the spectra
taken on 15-12-2007 ($-$16.5 \AA~) when compared 
with the observations on 22-01-2006 ($-$10 \AA~). The emission component in H$\beta$ absorption
line also increased, as shown in Fig. 3. 
Fe{\sc ii} 5169, 5235, 5316, 7513, 7712 \AA~and O{\sc i} 7772 \AA~appeared in 
emission when observed on 15-12-2007. This is also associated with an increase
in the emission strength of O{\sc i} 8446 \AA, Ca{\sc ii} triplet and Paschen
lines P11, P12, P14 and P17. 
The circumstellar disc has grown thick over a period of 23 months, which can
be deduced from the increase in intensity of low- and high-volt recombination lines. 

\subsection{H$\alpha$ equivalent width}
We have measured the H$\alpha$ EW for 150 CBe stars. 
The measured EW values are corrected for stellar
absorption using the theoretical values from the Kurucz database (Kurucz 1979). 
We have used the values corresponding to log(g) = 4.0, assuming the
candidates to be main-sequence stars. 
The corrected H$\alpha$ EWs are given in Table 1.  

The H$\alpha$ EW distribution of 150 CBe
stars is shown in Fig. 4 with a bin size of 10 \AA, 
which is higher than the measurement errors. 
The H$\alpha$ EW distribution of candidate CBe stars peak in the $-$10~$-$~$-$30
\AA~range, with 36 stars each in $-10$~$-$~$-$20~\AA~ and $-$20~$-$~$-$30~\AA~ bin. 
We found 21 CBe stars in $-$1~$-$~$-$10~\AA, 27 in $-$30~$-$~$-$40~\AA~ and 
16 stars in $-$40~$-$~$-$50~\AA~ EW
bins. There are 14 CBe stars whose H$\alpha$ EW values are less 
than $-$50~\AA~ with NGC 884(2) showing the extreme value of $-$70.4~\AA. 
About 80\% of our sample of CBe stars have H$\alpha$ EW in the range $-$1~$-$~$-$40
\AA, with 48\% in the range $-$10~$-$~$-$30 \AA.

\begin{figure}
\centerline{\includegraphics[width=9cm]{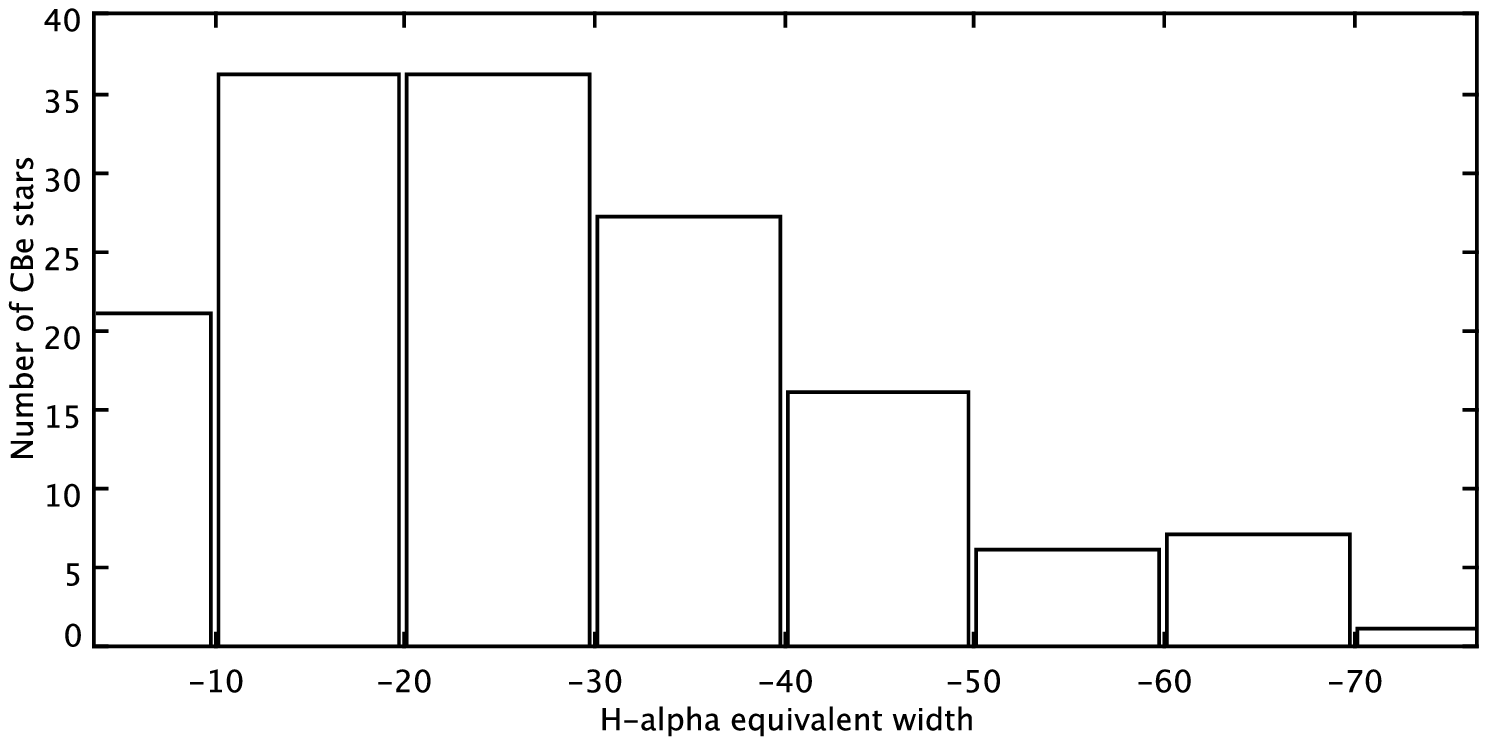}}
\caption{The distribution of H$\alpha$ EW of CBe stars.}
\end{figure}

\begin{figure}
\centerline{\includegraphics[width=9cm]{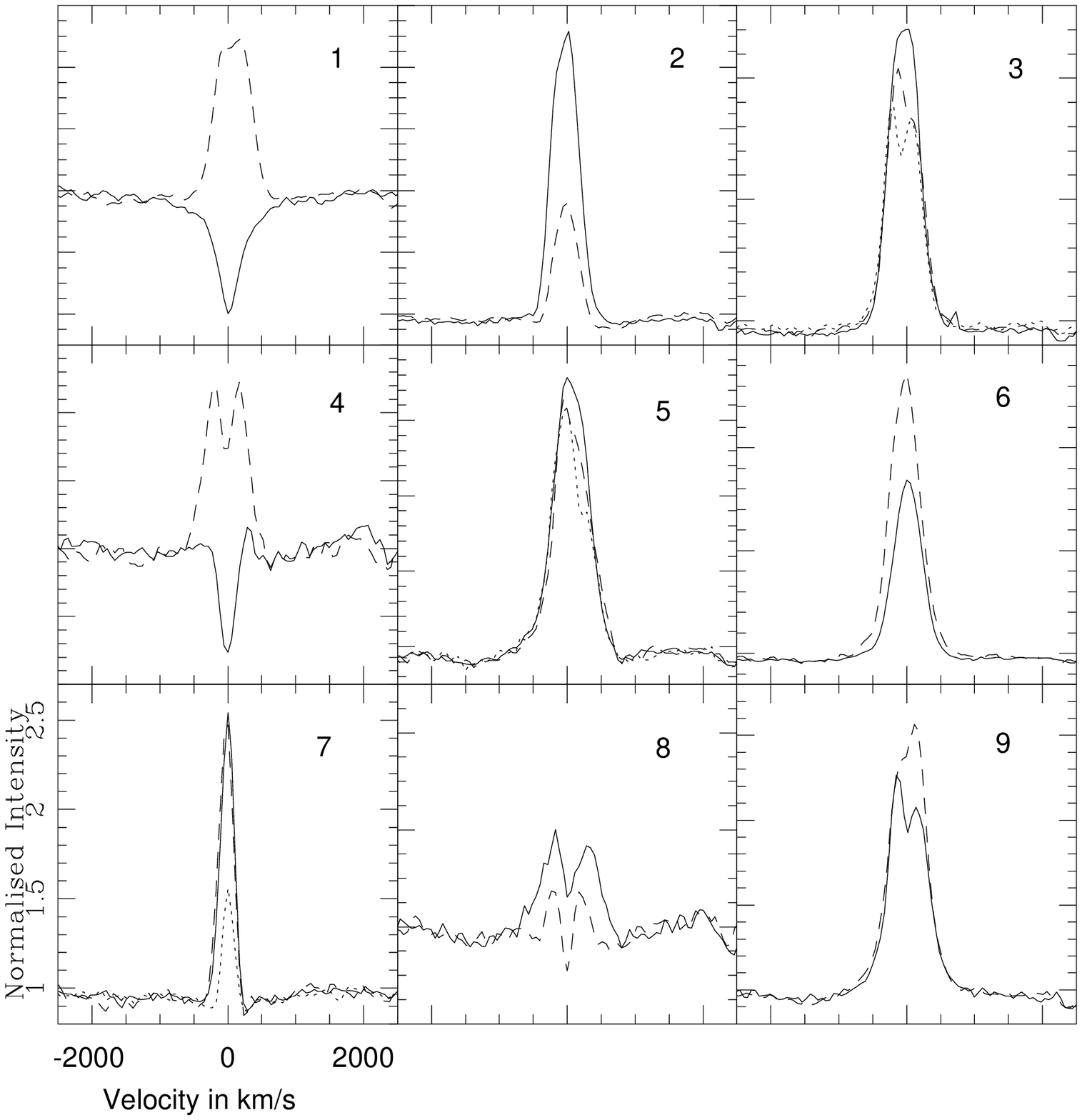}}
\caption{The variation in H$\alpha$ profile for e-stars (1) Berkeley 87(3), 
(2) NGC 659(2), (3) NGC 663(3), 
(4) NGC 663(13), (5) NGC 869(4), (6) NGC 884(1), (7) NGC 7419(H), (8) NGC 7419(K),
(9) NGC 7419(P) are shown. Initial observations are shown as solid lines 
followed by repeated observations in dashed and dotted lines respectively.}
\end{figure}

\subsubsection{Variability of H$\alpha$ profile}
McSwain et al.(2009) found 12 new transient CBe stars and confirm 17 additional CBe stars 
with relatively stable discs from H$\alpha$ spectroscopy of 296 stars in
eight open clusters. Of the total sample of 150 CBe stars in 39 open clusters 
we identified H$\alpha$ variability in 9 stars from multiple observations
over a period of a few years. A description of the profile
variability in these stars, which belong to 6 clusters, are listed below. 
In the present discussion we have not included the candidates which show slight change (EW of few \AA)
in emission strength.  

{\bf Berkeley 87(3)} : The observations of this star were made in 08-10-2005 and 25-10-2005. 
The profile changed from absorption to emission during this period, as shown
in Fig. 5.  

{\bf NGC 659(2)} : The profile changed from normal state when observed on 21-11-2005 to 
core-emission on 30-09-2006 with a significant reduction in emission strength. 

{\bf NGC 663(3)} : The emission profile was single peaked when observed on 07-10-2005 which changed to 
asymmetric emission on 09-10-2006. When observed on 24-10-2007 the profile was found to 
show a double-peaked feature.

{\bf NGC 663(13)} : The H$\alpha$ profile was in 
absorption when observed on 22-11-2005 which changed to a double-peaked emission
profile on 09-10-2006. This is a clear case of the formation of 
a circumstellar disc in a CBe star.

{\bf NGC 869(4)} : The profile changed from a symmetric emission profile when observed on 
22-01-2006 to an asymmetric profile on 24-10-2007. On subsequent observations
on 16-12-2007 the emission strength decreased.  

{\bf NGC 884(1)} : The emission strength of H$\alpha$ profile was enhanced when observed on 15-12-2007 
compared with that on 22-01-2006.

{\bf NGC 7419(H)} : The emission strength of H$\alpha$ decreased during
the observations on 15-07-2005, 08-08-2005 and 09-10-2006, where the 
profile shows a core-emission feature. 

{\bf NGC 7419(K)} : The profile resembled that of a typical CBe shell star, 
since the absorption component 
over the emission dips below the continuum level. 
The H$\alpha$ profile was double-peaked, with the violet part more intense than red
(V/R $>$ 1), when observed on 21-01-2006. The absorption 
component on emission deepened and fell below the continuum in the normalized spectra 
of 09-10-2006. 

{\bf NGC 7419(P)} : The profile was double-peaked when observed on 08-10-2005 
with the violet part of the 
double profile fading in intensity to a single profile on 10-10-2006.

The observed shape of H$\alpha$ emission line profile can be due to density
related effects in the circumstellar disc and is a function of viewing angle. 
Some of the candidates show double-peaked profile with differences in the
intensities of violet and red peaks (V $\neq$ R), which have been used
conventionally to understand the global oscillations in circumstellar discs 
(Mennickent, Sterken \& Vogt 1997). 
In the case of Berkeley 87(3), the circumstellar disc is formed during a 
short period ($\sim$17 days). 
The V/R ratio of NGC 663(P151) and NGC 869(6) were found to change 
over a period of time, suggesting the turbulent nature of the circumstellar
disc. From H$\alpha$ line profile analysis NGC 7419(K) is suspected to be a CBe
shell star.      

\begin{table} 
\caption{Spectral lines identified in the candidate CBe stars, with 
the number of emission and absorption profiles in brackets.}
\begin{tabular}{lrrrr}
\hline 
{\bf Fe{\sc ii}}        &               &                 &                 &             \\

4173(17e,7a)      & 4233(8e,4a)   & 4303(2e,2a)     & 4352(3a)      & 4385(2e)\\   
4417(2e,1a)       & 4515(6e,1a)   & 4520(8e,2a)     & 4523(3e,3a)   & 4549(7e,1a)\\
4556(5e,3a))      & 4584(30e,6a) & 4629(23e,1a)      & 4924(4e)     & 5018(51e,1a)\\
5169(65e,19a)     & 5198(23e,1a)  & 5235(34e,10a) & 5276(28e,7a)    & 5316(79e,7a) \\
5363(11e,1a)      & 5425(8e,1a)    & 5480(4e)      & 5496(1a)        & 5535(7e,1a)\\
5814(1a)         & 5957(1e)          & 5991(4e)      & 6084(1e)      & 6103(1a)\\
6148(12e,1a)    & 6248(9e)          & 6318(60e,4a)  & 6384(79e,3a)    & 6417(2e) \\
6432(1e)        & 6456(32e,7a)      & 6483(2e)      & 6516(35e)       & 7222(3e) \\
7308(1e)        & 7321(1e)          & 7462(8e)      & 7513(68e,2a)    & 7712(56e,1a) \\ 
                 &               &                   &                &  \\
{\bf O{\sc i} }        &               &                   &                &  \\
7772(71e,45a) & 8446(145e,3a) & &  & \\ 
                 &               &                   &                &  \\
{\bf Ca {\sc ii}}& & & &   \\
8498(103e,33a) & 8542(104e,39a) & 8662(98e,48a) & &  \\ 
                 &               &                   &                &  \\
{\bf Si {\sc ii}}& & & &    \\
6347(11e,28a) & 6371(6e,33a) & &  & \\ 
                 &               &                   &                &  \\
{\bf Mg {\sc ii}}& & & &    \\
7877(3e,1a) & 7896(8e,7a) & & & \\ 
\hline 
\end{tabular}
\end{table}

\subsection{Metallic lines}
Emission lines of metallic species, viz., Fe{\sc ii}, Ca{\sc ii}, 
O{\sc i}, Si{\sc ii}, Mg{\sc ii}, N{\sc ii}, form in the
circumstellar discs of Be stars which are not affected by the stellar
absorption features, unlike in the case of Balmer lines (Hanuschik 1987). 
Also, Fe{\sc ii} lines are least affected by broadening due to thermal effects
and Thomson scattering and hence they can be used to trace the density and velocity
structure of the envelope (Hanuschik 1988).  
The list of observed Fe{\sc ii}, Si{\sc ii}, Mg{\sc ii}, Ca{\sc ii}, O{\sc i} and 
N{\sc ii} lines of 150 CBe stars are shown in Table 3. 

We found that 131 (86\%) CBe stars show Fe{\sc ii} lines in their spectra. 
Among these, 92 have Fe{\sc ii} only in emission, while 5 have Fe{\sc ii} absorption lines. 
We identified 45 different Fe{\sc ii} spectral lines. 
Bochum 2(1) and NGC 146(S2) have only one Fe{\sc ii} 
absorption line while NGC 7128(2) has two and NGC 663(P151) has three lines. 
These stars have the least number of Fe{\sc ii} lines among the surveyed candidates.
On the other hand, NGC 884(2) has 25 Fe{\sc ii} emission lines and NGC 869(1)
shows 19 lines. The prominent Fe{\sc ii} lines present in the spectra are 
4584(36), 5018(52), 5169(84), 5316(86), 6318(64), 6384(82), 
7513(70) and 7712(57) ~\AA. 
The number of candidates which show these lines are given in brackets. 

The spectra of all but three stars in our sample show 8446 \AA~line. Only 116 stars have both 
7772 \AA~and 8446 \AA~oxygen lines in their spectra, 
where 71 stars (47\%) have both lines in emission. 
On the other hand, 36 stars (24\%) show only the 8446 \AA~oxygen line in their spectra. 

Either of the Si{\sc ii} lines 6347 \AA~and 6371 \AA~is seen in 32\% of the spectra in 
absorption or emission. 
For 24 stars these Si{\sc ii} lines are seen together either in emission or
absorption. 

The prominent Mg{\sc ii} lines in the spectra are of wavelengths 7877 \AA~and
7896 \AA, either of which are present in 20 stars. 
The N{\sc ii} 5005 \AA~line is present in emission in NGC 7419(C), 5463 \AA~in
NGC 6834(1), 5530 \AA~in NGC 7419(L), 5684 \AA~in NGC 7419(P) and 5942 \AA~in
NGC 7380(3). The presence of N{\sc ii} lines has to be confirmed through
repeated observations.  

The Ca{\sc ii} triplet (8498, 8542, 8662 \AA) is found to be blended with Paschen lines 
(P16, P15 and P13 respectively) in our sample of CBe stars. Ca{\sc ii} triplet is seen 
together either in absorption or emission in 130 stars. 
92 (60\%) stars show Ca{\sc ii} triplet in emission in their spectra. 
We found 100 stars to show Paschen 14 (8598 \AA) in emission 
while 144 have this line either in emission or absorption. 
About 117 candidates show more than 5 lines in Paschen series in their
spectra. 

\section{Conclusions}
1. We have presented the spectral details of 150 CBe stars of which 48 have been  
studied for the first time. This large data set
covers CBe stars of various spectral types and ages found in different cluster
environments of the northern open clusters. 
About 80\% of our sample of CBe stars have H$\alpha$ EW in the range $-$1~$-$~$-$40
\AA, with 48\% in the range $-$10~$-$~$-$30 \AA. 

2. Apart from the Balmer lines in emission, spectra of most of the stars show Fe{\sc ii}, 
Paschen and O{\sc i} lines in emission. 
About 86\% of the surveyed CBe stars show Fe{\sc ii} lines in their spectra. 
The prominent Fe{\sc ii} lines in our surveyed stars are 
4584, 5018, 5169, 5316, 6318, 6384, 7513 and 7712 \AA. 

3. We found long ($\sim$ years) and short (few days) term H$\alpha$ variability in 9 CBe stars which 
belong to 6 open clusters. In Berkeley 87(3) the profile is found to change from absorption to
emission in 17 days. For a few stars the V/R ratio changes over a period of 1
year. NGC 7419(K) is suspected to be a shell CBe star.

4. NGC 884(1) shows the presence of Fe{\sc ii}, O{\sc i}, Ca{\sc ii} triplet 
and Paschen emission lines in their spectra over a period of 23 months. 

5. The H$\alpha$ emission profile of the star NGC 663(13) changed from
absorption to a double-peaked profile, which is a clear case of the formation of
a circumstellar disc over a period of $\sim$1 year. 
This is accompanied by the formation of Ca{\sc ii} triplet absorption
lines, which trace the cooler part of the disc. 

\section{Acknowledgements}
We would like to acknowledge the help and support of the staff in CREST and Hanle
during the course of these observations. 
We would like to thank the anonymous referees for their suggestions which 
improved the quality of this paper. 
This research has made use of the WEBDA database, operated at the Institute for Astronomy 
of the University of Vienna.


\begin{thebibliography}{}
\bibitem{} Collins, G. W. II., 1987, in Proc. IAU Coll. 92, Physics of Be star, Cambridge
           University Press, Cambridge, p. 3
\bibitem{} Hanuschik R. W., 1987, A\&A, 173, 299
\bibitem{} Hanuschik R. W., 1988, A\&A, 190, 187
\bibitem{} Kurucz R. L., 1979, ApJS, 40, 1
\bibitem{} Malchenko S. L., Tarasov, A. E., 2008, in Choliy V. Ya., Ivashchenko G., 2008, eds, 
           Proc. YSC'15, p. 52
\bibitem{} Martayan C., Baade D., Fabregat J., 2009, IAUS, 256, 349
\bibitem{} Mathew B., Subramaniam A., Bhatt B. C., 2008, MNRAS, 388, 1879
\bibitem{} Mathew B., Subramaniam A., Bhavya B., 2010, BASI, 38, 35
\bibitem{} McSwain M. V., Gies. D. R., 2005, ApJS, 161, 118
\bibitem{} McSwain M. V., Huang W., Gies. D. R., Grundstrom E. D., Townsend R. H. D., 2008, ApJ, 672, 590
\bibitem{} McSwain M. V., Huang W., Gies. D. R., 2009, ApJ, 700, 1216 
\bibitem{} Mennickent R. E., Sterken C., Vogt N., 1997, A\&A, 326, 1167
\bibitem{} Pickles A. J., 1998, PASP, 110, 863
\bibitem{} Porter J. M., Rivinius T., 2003, PASP, 115, 1153
\bibitem{} Slettebak A., 1985, ApJS, 59, 769
\end{thebibliography}
\end{document}